\documentclass[prd,superscriptaddress,twocolumn,showpacs,amsmath,amssymb]{revtex4}
\usepackage{graphicx}
\usepackage{color}
\usepackage{dcolumn}
\usepackage{mathtools}
\usepackage{ulem}
\usepackage{color}

\usepackage{amsmath}
\usepackage{amssymb}
\usepackage{graphicx}
\usepackage{accents}
\usepackage{pstricks,pst-node,pst-coil,pst-plot}
\usepackage{mathtools}
\usepackage{ulem}
\usepackage{mathtools}
\usepackage{ulem}
\usepackage{graphicx}
\usepackage{color}
\usepackage{graphicx}
\usepackage{color}
\usepackage{dcolumn}
\usepackage{graphicx}
\usepackage{color}
\usepackage{dcolumn}
\usepackage{bm}
\usepackage{mathrsfs}
\usepackage{graphicx}
\usepackage{color}
\usepackage{dcolumn}
\usepackage{bm}
\usepackage{slashed}
\usepackage{epstopdf}
\usepackage{slashed}
\usepackage{bm}
\usepackage{slashed}
\usepackage{dcolumn}
\usepackage{bm}
\usepackage{slashed}
\usepackage{mathrsfs}
\usepackage{graphicx}
\usepackage{color}
\usepackage{dcolumn}
\usepackage{bm}
\usepackage{slashed}
\usepackage{epstopdf}
\usepackage{bm}
\usepackage{slashed}
\newcommand{\be}{\begin{equation}}
\newcommand{\ee}{\end{equation}}

\def\d{\mathrm{d}}
\def\p{\partial}
\usepackage{amsmath}\numberwithin{equation}{section}
\begin{document}
\title{On the angular momentum of compact binary coalescence}

\author{Xiaokai He}
\email{sjyhexiaokai@hnfnu.edu.cn}
 \affiliation{School of Mathematics and
Statistics, Hunan First Normal University, Changsha
410205, China}

\author{Xiaoning Wu}
\email{wuxn@amss.ac.cn}
\affiliation{Institute of Mathematics, Academy of Mathematics and Systems Science and Hua Loo-Keng Key Laboratory,
Chinese Academy of Sciences, Beijing 100190, China}

\author{Naqing Xie}
\email{nqxie@fudan.edu.cn}
 \affiliation{School of Mathematical Sciences, Fudan
University, Shanghai 200433, China}

\begin{abstract}
The supertranslation ambiguity issue of angular momentum is a long-standing problem in general relativity. Recently, there appeared the first definition of angular momentum at null infinity that is supertranslation invariant. However, in the compact binary coalescence community, supertranslation ambiguity is often ignored. This paper demonstrates that we have the happy circumstance that the newly defined angular momentum coincides with the classical definition at the quadrupole level.
\end{abstract}


\maketitle
\section{Introduction}
Gravitational radiation is an important prediction of general relativity (GR). Due to the complexity of diffeomorphisms, the existence of  gravitational waves (GW) became theoretically controversial since the genesis of GR \cite{Ken07}. Bondi, van der Burg, Metzner and Sachs established the  mathematical foundations of the theory of GW in the 1960s \cite{Bondi1962,Sachs1962}. The Bondi-Sachs framework describes the asymptotic structure and asymptotic symmetries near the future null infinity of asymptotic flat (AF) spacetimes. In their seminal works, Bondi \textit{et al.} successfully defined the mass of an AF spacetime and obtained the well-known Bondi mass loss formula. However, the asymptotic symmetric group is not the Poincar\'{e} group but an infinite dimensional analogue thereof, the Bondi-Metzner-Sachs (BMS) group. While the BMS group is structurally similar to the Poincar\'{e} group, the subgroup of translations is extended by a subgroup of supertranslations. The BMS group has a 4-dimensional Abelian canonical subgroup which is called the translation subgroup.
This allows the well-defined notions of the gravitational energy and the linear momentum and it ensures that the mass loss formula is invariant under supertranslations. Absence of the canonical $SO(3)$ subgroup causes the ambiguity in the definition of angular momentum and consequently the Bondi angular momentum flux is not supertranslation invariant.
The subtle supertranslation ambiguity of the angular momentum has been emphasized in the mathematical general relativity community over the past couple of decades \cite{Penrose1982}. There exist many candidates for the notions of the angular momentum in the literature \cite{Wini68,Bram75,Pri77,Str78,Ge81,Asht81,Asht82,Dr84,Dr85,Wa00} and unfortunately they are not supertranslation invariant. Recently, Po-Ning Chen, Mu-Tao Wang, Ye-Kai Wang and Shing-Tung Yau investigated the first definition of angular momentum  at null infinity that is supertranslation invariant \cite{CWY2021-2}. Different from the existing notions of the angular momentum, there is an important correction term in their newly defined expression. This additional term, which comes from solving the optimal isometric embedding equation for the Wang-Yau quasilocal mass \cite{WY09}, is closely related to the Bondi mass aspect and the closed potential of the Bondi shear tensor.

More and more gravitational events have been confirmed and a completely new window on the universe has been opened in the astronomical physics \cite{Abbott2016-1,Abbott2016-2,Abbott2017-1,Abbott2019,CCG2017,HJC2019}. In particular, on September 14, 2015 at 09:50:45 UTC, the two detectors of the Laser Interferometer Gravitational-wave Observatory (LIGO) observed the binary black hole merge \cite{Abbott2016-1}. Binary systems are among the most important sources of GW which are expected to be detected by the current or foreseeable gravitational wave detectors. The supertranslation ambiguity is often ignored in the compact binary coalescence (CBC) community in practice. The supermomentum contribution is small for the normally considered kick velocity. The detailed analysis was addressed in \cite{Asht20-1,Asht20-2}.

In this paper, we prove that, in the restricted context of compact binary coalescence, the newly defined angular momentum coincides with the classical definition at the quadrupole level. This result explicitly shows that we have the happy circumstance that the supertranslation ambiguity can be safely ignored in practice albeit it is conceptually important and mathematically difficult.

We always assume that the spacetime admits a conformal completion at null infinity \textit{\`{a} la} Penrose \cite{Pen65}. Throughout this paper, we make use of the convention that the speed of light and the universal gravitational constant are taken as fundamental units in a system. The signature of the spacetime metric is assumed to be $(-,+,+,+)$. Unless otherwise stated, Greek indices are used to label spacetime dimensions, lower case Latin indices are reserved for three-dimensional objects, and capital ones are for two-dimensional objects.

The paper is organized as follows. The notions of the total angular momentum, both the classical and the newly defined ones, are recalled in Section \ref{S2}. In Section \ref{S3}, we derive the coordinate transformation between harmonic coordinates and the Bondi coordinates, since the known results of the linear gravitational wave are formulated in the harmonic gauge while the notion of the angular momentum relies on the Bondi coordinates. In Section \ref{S4}, we prove that, in the linearised theory of GW for the compact binary coalescence, the correction term in the new expression indeed gives zero contribution in the flux integral at the quadrupole level. Conclusions and discussions are presented in the last section.

\section{Review of the notions of the angular momentum}\label{S2}

In this section, we give a quick review of the notions of the angular momentum. Within a Bondi-Sachs (BS) coordinate system $(u,r_{\rm BS},x^2,x^3)$, the physical spacetime metric takes the form
\be
\begin{split}
\d s^2=&-UV\d u^2-2U\d u\d r_{\rm BS}\\
&+r^2 h_{AB}(\d x^A+W^A\d u)(\d x^B+W^B\d u)
\end{split}
\ee
where $A,B=2,3$. The future null infinity, denoted by $\mathscr{I}^+$, corresponds to the null hypersurface $r_{\rm BS}=\infty$. The outgoing radiation condition implies the following expansions in inverse powers of $r_{\rm BS}$ \cite{CJK2002,CWY2021-2}:
\be
\begin{split}
h_{AB}=&q_{AB}+\frac{C_{AB}}{r_{\rm BS}}+\frac{1}{4r_{\rm BS}^2}|C|^2q_{AB}+O(r_{\rm BS}^{-3}),\\
W^{A}=&\frac{1}{2r_{\rm BS}^2}\nabla_BC^{AB}
+\frac{1}{r_{\rm BS}^3}\bigg(\frac{2}{3}N^A-\frac{1}{16}\nabla^A|C|^2\\
&-\frac{1}{2}C^{AB}\nabla^DC_{BD}\bigg)+O(r_{\rm BS}^{-4}),\\
U=&1-\frac{1}{16r_{\rm BS}^2}|C|^2+O(r_{\rm BS}^{-3}),\\
V=&1-\frac{2m}{r_{\rm BS}}+\frac{1}{r_{\rm BS}^2}\bigg(\frac{1}{3}\nabla^AN_A
+\frac{1}{4}\nabla^AC_{AB}\nabla_DC^{BD}\\
&+\frac{1}{16}
|C|^2\bigg)+O(r_{\rm BS}^{-3}).
\end{split}
\ee
Here $q_{AB}$ is the round metric on the unit sphere $S^2$, i.e.,
\be
q_{AB}dx^Adx^B=\d \theta^2+\sin^2\theta\d\varphi^2
\ee
and $\nabla_A$ denotes the covariant derivative associated with $q_{AB}$. The indices are contracted, raised and lowered with respect to $q_{AB}$. Moreover,
$m=m(u,x^A)$ is the mass aspect, $N_A=N_A(u,x^A)$ is the angular aspect and $C_{AB}=C_{AB}(u,x^A)$ is the shear tensor of this BS coordinates. The news tensor $N_{AB}$ is defined as
\be
N_{AB}=\partial_uC_{AB}.
\ee
It is well known that the evolution of the mass aspect function is
\be\label{masslossrate}
\partial_um=-\frac{1}{8}N_{AB}N^{AB}+
\frac{1}{4}\nabla^A\nabla^BN_{AB}.
\ee
The evolution of $N_A$ reads \cite{CJK2002,CWY2021-1}
\be
\begin{split}
\partial_uN_A=&\nabla_Am-\frac{1}{4}\nabla^D(\nabla_D\nabla^EC_{EA}
-\nabla_A\nabla^EC_{ED})\\
&+\frac{1}{4}\nabla_A(C_{BE}N^{BE})
-\frac{1}{4}\nabla_{B}(C^{BD}N_{DA})\\
&+\frac{1}{2}C_{AB}\nabla_DN^{DB}.
\end{split}
\ee

The standard formula for the BS energy-momentum at a $u$ cut along $\mathscr{I}^+$ are
\be
\begin{split}
E(u)&=\frac{1}{8\pi}\int_{S^2}2 m(u,\cdot),\\
P^k(u)&=\frac{1}{8\pi}\int_{S^2}2 m(u,\cdot)\tilde{X}^k,\ \ (k=1,2,3),
\end{split}
\ee
where
\be
\tilde{X}^k=(\sin\theta\cos\varphi,\sin\theta\sin\varphi,\cos\theta)
\ee
is the position vector of the standard unit sphere $S^2$ in $\mathbb{R}^3$.

A BMS field $Y$ is called a rotation BMS field if in a BS coordinate system,
\be
Y=\hat{Y}^A\frac{\partial}{\partial x^A}
\ee
where $\hat{Y}^A$ is a rotation Killing vector field on $S^2$. Precisely, the three rotation Killing vector fields on $S^2$ are
\be\label{K3}
\begin{split}
\hat{Y}_1&=\sin\varphi\frac{\partial}{\partial\theta}
+\cot\theta\cos\varphi\frac{\partial}{\partial\varphi},\\
\hat{Y}_2&=\cos\varphi\frac{\partial}{\partial\theta}-
\cot\theta\sin\varphi\frac{\partial}{\partial\varphi},\\
\hat{Y}_3&=\frac{\partial}{\partial\varphi}.
\end{split}
\ee

For a rotation BMS field $Y$ that is tangent to $u$ cuts on $\mathscr{I}^+$, the classical Bondi angular momentum of a  $u$ cut is defined as
\be\label{classicalAM}
\tilde{J}(u,Y)=\frac{1}{8\pi}\int_{S^2} Y^A\bigg(N_A-\frac{1}{4}C_{AB}\nabla_DC^{DB}\bigg).
\ee
This definition is that of Dray-Streubel in \cite{Dr84} and it does not satisfy the supertranslation invariance property. Very recently, Po-Ning Chen, Mu-Tao Wang, Ye-Kai Wang and Shing-Tung Yau investigated a new notion of the angular momentum that is supertranslation invariant \cite{CWY2021-2}. Aiming at eliminating the supertranslation ambiguity, they noticed that the shear tensor (as a symmetric traceless tensor) can be decomposed into
\be
C_{AB}=\nabla_A\nabla_Bc-\frac{1}{2}q_{AB}\Delta c
+\frac{1}{2}(\epsilon_A^{\ E}\nabla_E\nabla_B\underline{c}+\epsilon_B^{\ E}\nabla_E\nabla_A\underline{c})
\ee
where $\epsilon_{AB}$ is the volume form of $q_{AB}$. Here $c$ and $\underline{c}$ are the so-called closed and co-closed potentials of $C_{AB}$. The potentials  $c$ and $\underline{c}$ are uniquely determined by $C_{AB}$ if they are chosen to be of $l\geq 2$ harmonic modes.

For a rotation BMS field $Y$ that is tangent to $u$ cuts on $\mathscr{I}^+$, the new angular momentum of a  $u$ cut is defined as \cite{CWY2021-2}
\be
J(u,Y)=\frac{1}{8\pi}\int_{S^2} Y^A\bigg(N_A-\frac{1}{4}C_{AB}
\nabla_DC^{DB}-c\nabla_Am\bigg).
\ee
 Comparing with the classical angular momentum (\ref{classicalAM}), there appears an additional correction term in the new expression of the angular momentum
 \be\label{deltaj}
 \delta J(u,Y):=J(u,Y)-\tilde{J}(u,Y)=-\frac{1}{8\pi}\int_{S^2}cY^A\nabla_Am.
 \ee

\section{Bondi coordinates for the linear gravitational wave}\label{S3}

Consider a perturbation of the Minkowski spacetime in which the metric can be decomposed as
\be
g_{\mu\nu}=\eta_{\mu\nu}+h_{\mu\nu}, \ \ \ |h_{\mu\nu}|\ll 1.
\ee
 It is convenient to introduce the trace-reversed
perturbation $\bar{h}_{\mu\nu}$ by
\be
\bar{h}_{\mu\nu}:=h_{\mu\nu}-\frac{1}{2}\eta_{\mu\nu}h.
\ee
Up to the linear order in $h_{\mu\nu}$, the Einstein equation
\be
Ric(g)_{\mu\nu}-\frac{1}{2}Scal(g) g_{\mu\nu}=8\pi  T_{\mu\nu}
\ee
becomes
\be
\Box\bar{h}_{\mu\nu}
+\eta_{\mu\nu}\partial^{\rho}\partial^{\sigma}\bar{h}_{\rho\sigma}
-\partial^{\rho}\partial_{\nu}\bar{h}_{\mu\rho}
-\partial^{\rho}\partial_{\mu}\bar{h}_{\nu\rho}=-{16\pi }T_{\mu\nu}.
\ee
where $\Box=\eta^{\mu\nu}\partial_{\mu}\partial_{\nu}$ and the indices are raised and lowered by $\eta_{\mu\nu}.$

Under the harmonic gauge condition
\be
\partial^{\nu}\bar{h}_{\mu\nu}=0,
\ee
the linearised Einstein equation reads
\be\label{LiEH}
\Box \bar{h}_{\mu\nu}=-{16\pi }T_{\mu\nu}.
\ee
By using the Green function method, the linearised equation \eqref{LiEH} can be solved and the result is
 \be\label{linearsolution}
\bar{h}_{\mu\nu}(t,\vec{x})=
{4}\int\frac{T_{\mu\nu}(t-{|\vec{x}-\vec{x}'|},\vec{x}')}{|\vec{x}-\vec{x}'|}\d^3\vec{x}'.
\ee
At large distance from the compact source, we have the expansion
\be\label{hmunu1}\bar{h}_{\mu\nu}(t,\vec{x})=\frac{4}{r }
\int T_{\mu\nu}(t-r+{\vec{x}'\cdot\hat{n}},\vec{x}')
+O(\frac{1}{r^2}),
\ee
where $r=|\vec{x}|$ and $\hat{n}=\frac{\vec{x}}{r}.$

In terms of the momenta of the energy-momentum tensor $T_{\mu\nu}$,
\be
\begin{split}
S^{\mu\nu}(t)&:=\int T^{\mu\nu}(t,\vec{x})\d^3 x,\\
S^{\mu\nu,k}(t)&:=\int x^k T^{\mu\nu}(t,\vec{x})\d^3 x,\\
S^{\mu\nu,kl}(t)&:=\int x^k x^lT^{\mu\nu}(t,\vec{x})\d^3 x,\\
\end{split}
\ee
it follows from \eqref{hmunu1} that
\be\label{multipole-expansion}
\bar{h}^{\mu\nu}(t,\vec{x})
=\frac{4}{r}\bigg[S^{\mu\nu}+n_m\dot{S}^{\mu\nu,m}
+\frac{1}{2}n_mn_p\ddot{S}^{\mu\nu,mp}+\cdots\bigg]_{\rm ret}
\ee
where the subscript `ret' means the quantities are evaluated at the retarded time $t_R=t-{r}$ \cite{Maggiore2008}. Furthermore, via the linearised conservation law
\be
\partial_{\mu}T^{\mu\nu}=0,
\ee
one shows that the leading term of $\bar{h}_{ij}$ is
\be\label{hij}
\bar{h}_{ij}(t,\vec{x})
=\frac{2}{r}\ddot{M}^{ij}(t_R)
\ee
and
\be
M^{ij}=\int x^ix^j{T^{00}}\d ^3x
\ee
is the mass quadrupole of the source.

Now we consider a binary system with rest masses $M_1$ and $M_2$ and further assume that the relative coordinate is performing a circular motion.
By appropriately choosing the $(x,y,z)$ coordinates so that the orbit lies in the $(x,y)$ plane,  the relative motion $\vec{x}_0:=\vec{x}_1-\vec{x}_2$ is given by
\be\label{circular}
\begin{split}
x_0(t)&=R_0\cos(\omega_s t+\frac{\pi}{2}),\\
y_0(t)&=R_0\sin(\omega_s t+\frac{\pi}{2}),\\
z_0(t)&=0,
\end{split}
\ee
where $R_0$ is the distance between the two bodies and $\omega_s$ is the frequency of the motion. It is well known that, in the center-of-mass (CM) frame, the mass quadrupole of a binary system is given by \cite{Maggiore2008}
\be\label{Mij}
M^{ij}(t)=\mu x_0^ix_0^j
\ee
where $\mu=\frac{M_1M_2}{M_1+M_2}$ is the reduced mass of the binary system. Combining \eqref{hij}, \eqref{circular} and \eqref{Mij}, one finds that the nonzero components of $\bar{h}_{ij}$ are
\be\label{hbarij}
\begin{split}
\bar{h}_{11}(t,\vec{x})&=-\bar{h}_{22}(t,\vec{x})=\frac{8\mu R_0^2\omega_s^2}{r}\cos2\omega_st_R,\\
\bar{h}_{12}(t,\vec{x})&=\bar{h}_{21}(t,\vec{x})=\frac{8\mu R_0^2\omega_s^2}{r}\sin 2\omega_st_R.
\end{split}
\ee

Moreover, the leading order term of  $\bar{h}_{00}$ and $\bar{h}_{0i}$ can be computed from
\eqref{multipole-expansion} and the results are
\be\nonumber
\begin{split}
\bar{h}_{00}&=\frac{4}{r}(M+\frac{1}{2}\ddot{M}_{ij}n^in^j),\\
\bar{h}_{0i}&=-\frac{2}{r}\ddot{M}_{ij}n^j.
\end{split}
\ee
More explicitly,
\be\label{h0mu}
\begin{split}
\bar{h}_{00}&=\frac{4}{r}(M+F),\\
 \bar{h}_{01}&=-\frac{8\mu R_0^2\omega_s^2}{r}\cos(2\omega_st_R-\varphi)\sin\theta,\\
  \bar{h}_{02}&=-\frac{8\mu R_0^2\omega_s^2}{r}\sin(2\omega_st_R-\varphi)\sin\theta,\\
  \bar{h}_{03}&=0,
\end{split}
\ee
with $$M=M_1+M_2$$ and
\be\nonumber
\begin{split}
F&=\frac{1}{2}\ddot{M}_{ij}n^in^j\\
&=\mu R_0^2\omega^2_s\bigg[\cos(2\omega_st_R-2\varphi)-
\frac{1}{2}\cos(2\omega_st_R-2\theta-2\varphi)\\
&-\frac{1}{2}\cos(2\omega_st_R+2\theta-2\varphi)\bigg].
\end{split}
\ee

Eqs. \eqref{hbarij} and \eqref{h0mu} give the form of the metric as
\begin{widetext}
{{
\be\label{318}
g_{\mu\nu}^{( H)}=\left(
    \begin{array}{cccc}
      -1+\frac{2M}{r}+\frac{2F}{r} & -\frac{8\mu R_0^2\omega_s^2\cos(2\omega_st_R-\varphi)\sin\theta}{r} & -\frac{8\mu R_0^2\omega_s^2\sin(2\omega_st_R-\varphi)\sin\theta}{r}& 0 \\
       -\frac{8\mu R_0^2\omega_s^2\cos(2\omega_st_R-\varphi)\sin\theta}{r}& 1+\frac{8\mu R_0^2\omega_s^2\cos2\omega_st_R}{r}+ \frac{2M}{r}+\frac{2F}{r} & \frac{8\mu R_0^2\omega_s^2\sin 2\omega_st_R}{r} & 0 \\
       -\frac{8\mu R_0^2\omega_s^2\sin(2\omega_st_R-\varphi)\sin\theta}{r}& \frac{8\mu R_0^2\omega_s^2\sin 2\omega_st_R}{r} &1 -\frac{8\mu R_0^2\omega_s^2\cos2\omega_st_R}{r}+ \frac{2M}{r}+\frac{2F}{r} & 0 \\
      0 & 0 & 0 &  1+\frac{2M}{r} +\frac{2F}{r}\\
    \end{array}
  \right).
\ee}}
\end{widetext}
Here the script `(H)' means that the metric is expressed in the harmonic coordinates.
To calculate the correction term $\delta J$ in the new angular momentum expression, the above metric form should be transformed  into certain Bondi-Sachs (BS) form. In a BS form,
\be
\begin{split}
g_{(\rm BS)}^{uu}&=0,\ \ g_{(\rm BS)}^{uA}=0,\\
g^{ur}_{(\rm BS)}&=-1+O(\frac{1}{r_{\rm BS}^2}),\\
g^{rr}_{(\rm BS)}&=1-\frac{2m}{r_{\rm BS}}+O(\frac{1}{r_{\rm BS}^2}),\\
g^{rA}_{(\rm BS)}&=\frac{1}{2 r_{\rm BS}^2}\nabla_BC^{AB}+O(\frac{1}{r_{\rm BS}^3}),\\
g^{AB}_{(\rm BS)}&=\frac{1}{r_{\rm BS}^2}q^{AB}
-\frac{1}{r_{\rm BS}^3}C^{AB}+O(\frac{1}{r_{\rm BS}^4}).
\end{split}
\ee

The coordinate transformation can be done in the following way. Within the harmonic coordinate system, in general,  we have
\be\label{320}
\begin{split}
g^{(H)}_{00}
=&-1+\frac{2M}{r}+\frac{1}{r}H_{00}(t-r,\theta,\varphi),\\
g^{(H)}_{0i}=&
\frac{1}{r}H_{0i}(t-r,\theta,\varphi),\\
g_{ij}^{(H)}=&\delta_{ij}+\frac{2M}{r}\delta_{ij}
+\frac{1}{r}H_{ij}(t-r,\theta,\varphi).
\end{split}
\ee
For the discussion of gravitational radiation, we only need to work up to the order of $\frac{1}{r}$ in the harmonic coordinates and determine the corresponding BS coordinates such that the Bondi gauge is imposed up to the order of $\frac{1}{r_{B}}$ beyond the leading order metric. To perform the coordinate transformation, it is more convenient to work with the components of the inverse metric. In the harmonic gauge, direct calculation shows that the components of the inverse metric are given by
\be\label{harmonic-inverse}
\begin{split}
g^{00}_{(H)}=&-1-\frac{2M}{r}-\frac{1}{r}H^{00}
+O(\frac{1}{r^2}),\\
g^{0i}_{(H)}=&-\frac{1}{r}H^{0i}+O(\frac{1}{r^2}),\\
g^{ij}_{(H)}=&(1-\frac{2M}{r})\delta_{ij}
-\frac{H^{ij}}{r}+O(\frac{1}{r^2}).
\end{split}
\ee
where
\be
H^{\mu\nu}=\eta^{\mu\alpha}\eta^{\nu\beta}H_{\alpha\beta}.
\ee
 The coordinate transformations from the harmonic gauge to the BS gauge are split into two steps. Firstly, we imposes the gauge conditions on the inverse Bondi metric so that $g^{uu}_{({\rm BS})}$ and $g^{uA}_{({\rm BS})}$ vanish with the accuracy of order $\frac{1}{r_{\rm BS}}$. It also makes $r_{\rm BS}$ an areal radius and  relates the BS coordinate $\tilde{u}$ to the harmonic gauge retarded time $t-r$.

Let
\be\label{preBScoordinate}
\begin{split}
\tilde{u}=&t-{r}-{2M}\ln\frac{r}{r_0},\\
\tilde{r}=&r+{M},\\
\tilde{\theta}=&\theta=\arccos\frac{z}{r},\\
\tilde{\varphi}=&\varphi=\arctan\frac{y}{x}.
\end{split}
\ee
The coordinates $\{\tilde{u},\tilde{r},\tilde{\theta},\tilde{\varphi}\}$ are not precisely the BS coordinates up to the order of $1/\tilde{r}$ since the metric does not satisfy all the requirements of the Bondi gauge \cite{Tahura2021}. We will thus denote by $g^{\mu\nu}_{(I)}$ this `intermediate' metric. More precisely, $g^{\mu\nu}_{(I)}$ can be computed from the harmonic gauge metric (\ref{harmonic-inverse}) using the transformation law
\be
g^{\mu\nu}_{(I)}=g^{\alpha\beta}_{(H)}\frac{\p\tilde{x}^{\mu}}{\p x^{\alpha}}\frac{\p\tilde{x}^{\nu}}{\p x^{\beta}}.
\ee
Direct calculation gives
\be
g^{00}_{(I)}=-\frac{1}{r}[H_{00}+2H_{0i}n^i+H_{ij}n^in^j]+O(\frac{1}{r^2}).
\ee
For the metric \eqref{318}, one can read off $H_{00}$, $H_{0i}$ and $H_{ij}$ from \eqref{320} and this immediately yields $H_{00}+2H_{0i}n^i+H_{ij}n^in^j=0$.

In general, recall that the harmonic gauge condition gives \cite{Tahura2021}
\be
H_{00}+H=n^in^jH_{ij},
\ee
and
\be
H_{0i}=-n^jH_{ij}+\frac{H}{2}n^i.
\ee
Then $g^{00}_{(I)}$ can be simplified as
\be
\begin{split}
g^{00}_{(I)}&=-\frac{1}{r}[H_{00}+2(-n^in^jH_{ij}+\frac{H}{2})
+H_{ij}n^in^j]+O(\frac{1}{r^2})\\
&=-\frac{1}{r}[H_{00}-n^in^jH_{ij}+H]+O(\frac{1}{r^2})\\
&=O(\frac{1}{r^2})=O(\frac{1}{\tilde{r}^2}),
\end{split}
\ee
which implies that the Bondi gauge condition ${g^{uu}=0}$ is satisfied at the required order.
Similarly, we have
\be
\begin{split}
g^{0A}_{(I)}&=O(\frac{1}{\tilde{r}^3}),\\  g^{01}_{(I)}&=-1+{\frac{1}{\tilde{r}}
\frac{H}{2}}+O(\frac{1}{\tilde{r}^2}),\\
g^{11}_{(I)}&=1-\frac{1}{\tilde{r}}(2M+{H_{ij}n^in^j})
+O(\frac{1}{\tilde{r}^2}),\\
g^{1A}_{(I)}&=\frac{1}{\tilde{r}^2}{H_{0i}\p^An^i}
+O(\frac{1}{\tilde{r}^3}),\\
g^{AB}_{(I)}&=\frac{1}{\tilde{r}^2}q^{AB}-{\frac{1}{\tilde{r}^3}H_{ij}(\p^An^i)
(\p^Bn^j)}+O(\frac{1}{\tilde{r}^4}).
\end{split}
\ee
Here $\partial^A=q^{AB}\partial_B$ denotes the derivative on the sphere.

So far, the metric in the $\{\tilde{u},\tilde{r},\tilde{\theta},\tilde{\varphi}\}$ coordinate system looks similar to a BS form up to order $\frac{1}{r}$. The metric can be further put into the Bondi gauge in a perturbative manner. For this purpose,
we parameterise the perturbative coordinate transformation in terms of a vector $\xi^{\mu}$,
\be
x^{\mu}_{({\rm BS})}=\tilde{x}^{\mu}+\xi^{\mu}(\tilde{x}).
\ee
This coordinate transformation will bring the metric $g^{\mu\nu}_{(I)}$ to the Bondi-Sachs form $g^{\mu\nu}_{({\rm BS})}$ via the transformation
\be
g^{\mu\nu}_{({\rm BS})}=g^{\alpha\beta}_{(I)}\frac{\p x^{\mu}_B}{\p\tilde{x}^{\alpha}}\frac{\p x^{\nu}_B}{\p\tilde{x}^{\beta}}
=g^{\alpha\beta}_{(I)}\bigg(\delta^{\mu}_{\alpha}+
\frac{\p\xi^{\mu}}{\p\tilde{x}^{\alpha}}\bigg)
\bigg(\delta^{\nu}_{\beta}+
\frac{\p\xi^{\nu}}{\p\tilde{x}^{\beta}}\bigg).
\ee
To seek the required perturbative gauge $\xi^{\mu}$, we set
\be
\xi^{\mu}=\bigg(\frac{\xi^0_{(1)}(\tilde{u},
\tilde{\theta},\tilde{\varphi})}{\tilde{r}},
\frac{\xi^1_{(1)}(\tilde{u},
\tilde{\theta},\tilde{\varphi})}{\tilde{r}},
\frac{\xi^2_{(1)}(\tilde{u},
\tilde{\theta},\tilde{\varphi})}{\tilde{r}^2},
\frac{\xi^3_{(1)}(\tilde{u},
\tilde{\theta},\tilde{\varphi})}{\tilde{r}^2}\bigg).
\ee
Equivalently,
\be
\begin{split}
u_B&=\tilde{u}+\frac{\xi^0_{(1)}(\tilde{u},
\tilde{\theta},\tilde{\varphi})}{\tilde{r}},\\
r_B&=\tilde{r}+\frac{\xi^1_{(1)}(\tilde{u},
\tilde{\theta},\tilde{\varphi})}{\tilde{r}},\\
\theta_B&=\tilde{\theta}+\frac{\xi^2_{(1)}(\tilde{u},
\tilde{\theta},\tilde{\varphi})}{\tilde{r}^2},\\
\varphi_B&=\tilde{\varphi}+\frac{\xi^2_{(1)}(\tilde{u},
\tilde{\theta},\tilde{\varphi})}{\tilde{r}^2}.
\end{split}
\ee
Straightforward calculation shows
\be
\begin{split}
g^{00}_{({\rm BS})}&=O(\frac{1}{r_{\rm BS}^2}),\ \
g^{0A}_{({\rm BS})}=O(\frac{1}{r_{\rm BS}^3}),\\
g^{01}_{({\rm BS})}&=-1+\frac{1}{r_{\rm BS}}\bigg[\frac{H}{2}-
\frac{\p\xi^0_{(1)}}{\p\tilde{u}}\bigg]+O(\frac{1}{r_{\rm BS}^2}),\\
g^{11}_{({\rm BS})}&=1-\frac{1}{r_{\rm BS}}\bigg[2M+H_{ij}n^in^j+
2\frac{\p\xi^1_{(1)}}{\p\tilde{u}}\bigg]
+O(\frac{1}{r_{\rm BS}^2}),\\
g^{12}_{({\rm BS})}&=\frac{1}{r_{\rm BS}^2}\bigg[-H_{ij}n^i\frac{\p n^j}{\p\theta}-
\frac{\p\xi^2_{(1)}}{\p\tilde{u}}\bigg]+O(\frac{1}{r_{\rm BS}^3}),\\
g^{13}_{({\rm BS})}&=\frac{1}{r_{\rm BS}^2}\bigg[-\frac{1}{\sin^2\theta}H_{ij}n^i\frac{\p n^j}{\p\varphi}-
\frac{\p\xi^3_{(1)}}{\p\tilde{u}}\bigg]+O(\frac{1}{r_{\rm BS}^3}),\\
g^{AB}_{({\rm BS})}&=\frac{1}{r_{\rm BS}^2}(\p^An^i)(\p^Bn^i)-\frac{1}{r_{\rm BS}^3}H_{ij}(\p^An^i)
(\p^Bn^j)+O(\frac{1}{r_{\rm BS}^4})\\
&=\frac{q_{AB}}{r_{\rm BS}^2}-\frac{1}{r_{\rm BS}^3}H_{ij}(\p^An^i)
(\p^Bn^j)+O(\frac{1}{r_{\rm BS}^4}).
\end{split}
\ee

From the above results, we find that the {Bondi shear} is
\be
{C^{AB}=H_{ij}(\p^An^i)
(\p^Bn^j)}.
\ee
To achieve the Bondi gauge,  we should choose $\xi_{(1)}^0,
\xi_{(1)}^A$ such that
\be
\frac{H}{2}-\frac{\p\xi^0_{(1)}}{\p\tilde{u}}=0,
\ee
and
\be
-H_{ij}n^i\p^An^j-\frac{\p\xi^A_{(1)}}{\p\tilde{u}}
=\frac{1}{2}\nabla_BC^{AB}.
\ee
Moreover, one requires that $\frac{\partial\xi^{1}_{(1)}}{\partial \tilde{u}}=\frac{1}{2}H_{ij}n^in^j$ to preserve \eqref{masslossrate} and the Bondi mass aspect now becomes
\be
m={M+H_{ij}n^in^j}.
\ee

By using the harmonic gauge condition again, it follows that the Bondi shear tensor $C^{AB}$ is only related to the transverse-traceless part of $H_{ij}$ as \cite{Tahura2021}
\be\label{C1}
{C^{AB}=H^{\rm TT}_{ij}(\p^An^i)
(\p^Bn^j)},
\ee
where
\be\label{HT1}
H^{\rm TT}_{ij}=\Lambda_{ijkl}H^{kl}
\ee
with
\be
\begin{split}
\Lambda_{ijkl}&=P_{ik}P_{jl}-\frac{1}{2}P_{ij}P_{kl},\\
P_{ij}&=\delta_{ij}-n_in_j.
\end{split}
\ee
Recently, Blanchet \textit{et al.} transformed the metric of an isolated matter source in the multipolar post-Minkowskian (MPM) approximation from the harmonic coordinates to the Newman-Unti coordinates (which are equivalent to the Bondi-Sachs coordinates). They also obtained the mass and  angular momentum aspects, as well as the Bondi shear in terms of the canonical multipole moments
\cite{Blanchet2021}. Up to the linear order of $G$,
 the Bondi shear tensor $C^{AB}$ reads
\be
{C^{AB}=H^{\rm TT}_{ij}(\p^An^i)
(\p^Bn^j)},
\ee
and the Bondi mass aspect is
\be
m=\sum_{l=0}^{+\infty}\frac{(l+1)(l+2)}{2l!}
n_LM_L^{(l)},
\ee
which reduces to
\be
m={M+3\ddot{M}_{ij}n^in^j}
\ee
at the quadrupole level.

\section{Correction term $\delta J$ for linear gravitational wave}{\label{S4}}

Now we are in the position to calculate the contribution of the additional correction term $\delta J$ \eqref{deltaj} in the new angular momentum expression.

Recall that the closed potential $c$ and co-closed potential $\underline{c}$ of $C_{AB}$ are related with $C_{AB}$ via
\be\label{cab}
C_{AB}=\nabla_A\nabla_Bc-\frac{1}{2}q_{AB}\Delta c
+\frac{1}{2}(\epsilon_A^{\ E}\nabla_E\nabla_B\underline{c}+\epsilon_B^{\ E}\nabla_E\nabla_A\underline{c})
\ee

Then
\be\label{cequation}
\begin{split}
&\nabla^A\nabla^BC_{AB}\\
=&\nabla^A\nabla^B\bigg[\nabla_A\nabla_Bc-\frac{1}{2}q_{AB}\Delta c+\frac{1}{2}(\epsilon_A^{\ E}\nabla_E\nabla_B\underline{c}\\
&+\epsilon_B^{\ E}\nabla_E\nabla_A\underline{c})\bigg]\\
=&\nabla^A\nabla^B\nabla_A\nabla_Bc-\frac{1}{2}\Delta^2c
+\frac{1}{2}\epsilon_A^{\  E}\nabla^A\nabla^B\nabla_B\nabla_E\underline{c}\\
&+\frac{1}{2}\epsilon_B^{\  E}\nabla^A\nabla^B\nabla_A\nabla_E\underline{c}\\
=&\nabla^A[\nabla_A\nabla^B\nabla_Bc+R_A^{\ E}\nabla_Ec]
+\frac{1}{2}\epsilon_A^{\ E}\nabla^A
\nabla^B\nabla_E\nabla_B\underline{c}\\
&-\frac{1}{2}\Delta^2c+\frac{1}{2}\epsilon_B^{\ E}\nabla^A(\nabla_A\nabla^B\nabla_E\underline{c}
+R^{B\ \ \ F}_{\ AE}\nabla_F\underline{c})\\
=&\frac{1}{2}\Delta^2c+\nabla^A(\delta_A^E\nabla_Ec)
+\frac{1}{2}\epsilon_A^{\ E}\nabla^A(\nabla_E\nabla^B\nabla_B\underline{c}
\\&+R^{ \ F}_{E}\nabla_F\underline{c})+\frac{1}{2}\epsilon_B^{\ E}\nabla^A(\nabla_A\nabla^B\nabla_E\underline{c}
R^{B\ \ \ F}_{\ EB}\nabla_F\underline{c})\\
=&\frac{1}{2}\Delta^2c+\Delta c
+\frac{1}{2}\epsilon_A^{\ E}\nabla^A(\delta_E^{\ F}\nabla_F\underline{c})\\
&+\frac{1}{2}\epsilon_{B}^{\ E}\nabla^A(R^{B\ \ \ F}_{\ EB}\nabla_F\underline{c})\\
=&\frac{1}{2}\Delta^2c+\Delta c+\frac{1}{2}\epsilon_{B}^{\ E}\nabla^A(\epsilon_A^{\ B}\epsilon_E^{\ F}\nabla_F\underline{c})\\
=&\frac{1}{2}\Delta^2c+\Delta c,
\end{split}
\ee
where we have used the fact that, for the round 2-sphere $(S^2,q_{AB})$,
\be
R_{ABCD}=\epsilon_{AB}\epsilon_{CD},\ R_{A}^B=\delta^B_A.
\ee

To obtain the closed potential $c$ from $C_{AB}$, we  expand $\nabla^A\nabla^BC_{AB}$ and $c$ in terms of the spherical harmonics
\be\label{cexpand}
\begin{split}
\nabla^A\nabla^BC_{AB}&=\sum_{l\geq2}\sum_{m=-l}^{l}f_{lm}Y_{lm}(\theta,\varphi),\\
c&=\sum_{l\geq 2}\sum_{m=-l}^{l}c_{lm}Y_{lm}(\theta,\varphi).
\end{split}
\ee
Inserting \eqref{cexpand} into \eqref{cequation}, one has
\be\label{clm}
c_{lm}=\frac{2f_{lm}}{(l-1)l(l+1)(l+2)},\ \ (l\geq 2).
\ee

At the quadrupole level, $H_{ij}=2\ddot{M}_{ij}+2F\delta_{ij}$ and it follows from \eqref{C1} and \eqref{HT1} that
\be
C_{AB}=\left(
    \begin{array}{cc}
      H_+ & H_{\times}\sin\theta    \\
      H_{\times}\sin\theta & - H_+\sin^2\theta   \\
    \end{array}
  \right)
\ee
where
\be
\begin{split}
H_+=&\ddot{M}_{11}(\cos^2\theta\cos^2\varphi-\sin^2\varphi)
+\ddot{M}_{12}(1+\cos^2\theta)\sin2\varphi\\
&-\ddot{M}_{13}\sin2\theta\cos\varphi -\ddot{M}_{22}(\cos^2\varphi-\sin^2\varphi\cos^2\theta)\\
&
-\ddot{M}_{23}\sin2\theta\sin\varphi
+\ddot{M}_{33}\sin^2\theta,\\
H_{\times}=&(\ddot{M}_{22}-\ddot{M}_{11})\cos\theta\sin2\varphi
+2\ddot{M}_{12}\cos\theta\cos2\varphi\\
&+2\ddot{M}_{13}\sin\theta\sin\varphi-
2\ddot{M}_{23}\sin\theta\cos\varphi,
\end{split}
\ee

Further, we have
\be
\begin{split}
&\nabla^A\nabla^BC_{AB}\\
=&-\ddot{M}_{11}-\ddot{M}_{22}
+2\ddot{M}_{33}-3(\ddot{M}_{11}+\ddot{M}_{22}-2\ddot{M}_{33})
\cos2\theta\\
&-\frac{3}{2}(\ddot{M}_{11}-\ddot{M}_{22})\cos(2\theta-2\varphi)
+6\ddot{M}_{23}\cos(2\theta-\varphi)\\
&+
3\ddot{M}_{11}\cos2\varphi-3\ddot{M}_{22}
\cos2\varphi-\frac{3}{2}\ddot{M}_{11}
\cos(2\theta+2\varphi)\\
&+\frac{3}{2}\ddot{M}_{22}\cos(2\theta+2\varphi)
-6\ddot{M}_{23}\cos(2\theta+\varphi)\\
&+3\ddot{M}_{12}
\sin(2\theta-2\varphi)+6\ddot{M}_{13}\sin(2\theta-\varphi)
+6\ddot{M}_{12}\sin2\varphi\\
&-3\ddot{M}_{12}
\sin(2\theta+2\varphi)
+6\ddot{M}_{13}\sin(2\theta+\varphi),
\end{split}
\ee
which yields
\be
\begin{split}
c=&\frac{-1}{12}\bigg[\ddot{M}_{11}+\ddot{M}_{22}
-2\ddot{M}_{33}+3(\ddot{M}_{11}+\ddot{M}_{22}-2\ddot{M}_{33})
\cos2\theta\\
&+\frac{3}{2}(\ddot{M}_{11}-\ddot{M}_{22})\cos(2\theta-2\varphi)
-6\ddot{M}_{23}\cos(2\theta-\varphi)\\
&-
3\ddot{M}_{11}\cos2\varphi+3\ddot{M}_{22}
\cos2\varphi+\frac{3}{2}\ddot{M}_{11}
\cos(2\theta+2\varphi)\\
&-\frac{3}{2}\ddot{M}_{22}\cos(2\theta+2\varphi)+6\ddot{M}_{23}
\cos(2\theta+\varphi)\\
&-3\ddot{M}_{12}
\sin(2\theta-2\varphi)-6\ddot{M}_{13}\sin(2\theta-\varphi)
-6\ddot{M}_{12}\sin2\varphi\\
&+3\ddot{M}_{12}
\sin(2\theta+2\varphi)-6\ddot{M}_{13}
\sin(2\theta+\varphi)\bigg].
\end{split}
\ee

Moreover, at the quadrupole level,  the Bondi mass aspect is
\be
\begin{split}
m=&M+3n^in^j\ddot{M}_{ij}\\
=&M_1+M_2+3\big(\ddot{M}_{11}\sin^2\theta\cos^2\varphi
+\ddot{M}_{12}\sin^2\theta\sin 2\varphi\\
&+
\ddot{M}_{13}\sin2\theta\cos\varphi +
\ddot{M}_{22}\sin^2\theta\sin^2\varphi\\
&+
\ddot{M}_{23}\sin2\theta\sin\varphi
+\ddot{M}_{33}\cos^2\theta\big).
\end{split}
\ee
By direct calculation, one finally finds that the integrals involving the correction terms in the new angular momentum expression are
\be
\begin{split}
&\int_{S^2} cY_1^A\nabla_Am=0,\\
&\int_{S^2} cY_2^A\nabla_Am=0,\\
 &\int_{S^2} cY_3^A\nabla_Am=0,
\end{split}
\ee
where $Y_1^A,Y_2^A$ and $Y_3^A$ are the three rotation Killing vector fields \eqref{K3} on the round sphere $S^2$.

\section{Conclusions and Discussions}\label{S5}
The notion of the angular momentum can be shifted by supertranslations. Such ambiguity has been an obstacle to seek a satisfactory notion of the angular momentum. Recently, Po-Ning Chen, Mu-Tao Wang, Ye-Kai Wang and Shing-Tung Yau investigated the first definition of the angular momentum that is supertranslation invariant in the full nonlinear theory of GR. On the other side, in the compact binary coalescence community, supertranslation ambiguity is often ignored in practice. Under certain assumptions normally made in the CBC literature, the orders of the magnitude of the ambiguity contribution are smaller than the statistical errors of GW detectors \cite{Asht20-1}. In this paper, we investigate the supertranslation ambiguity issue in the restricted context of compact binary coalescence. We have shown that, in the linearised theory of GW, the new angular momentum coincides with the classical definition at the quadrupole level.

However, there is a caveat in this current work. Within the frame of the linearised theory of GW, we have not checked the contribution of the additional correction term in the full nonlinear theory. The main difficulty we have encountered comes from the lack of fully explicit expressions of the radiative metrics (in terms of the BS coordinates). We believe that the recently proposed angular momentum is a very nice definition and it should deserve much more attention and investigation in the future. In particular, one wishes to seek the physical origin and significance of the newly added correction term. There is a question for the geometric analysis community: Can one provide a concrete physical example of radiative spacetime in which the correction term has nonzero contribution?

\section*{Acknowledgements}
X. He was partially supported by the Key Project of Education Department of Hunan Province (21A0576). X. Wu was partially supported by the National Natural Science Foundation of China (11731001). N. Xie was partially supported by the National Natural Science Foundation of China (11671089).

\end{document}